\documentclass{ar-1col}
\usepackage[numbers]{natbib}
\usepackage{url}
\usepackage{amsmath}
\usepackage{natbib}
\usepackage{isotope}
\usepackage{xspace}

\newcommand{\Mhi}{M_{\text{hi}}}

\newcommand{\adhoc}{\textit{ad hoc}\xspace}
\newcommand{\apriori}{\textit{a priori}\xspace}

\jname{Xxxx. Xxx. Xxx. Xxx.}
\jvol{AA}
\jyear{YYYY}
\doi{10.1146/((please add article doi))}

\begin{document}

\markboth{Crawford \textit{et al.}}{A Vision for the Science of Rare Isotopes}

\title{A Vision for the Science of Rare Isotopes}

\author{
H.~L.~Crawford$^{1}$, 
K.~Fossez$^{2,3}$, 
S.~K\"onig$^4$ and 
A.~Spyrou$^{5,6}$
\affil{$^1$Nuclear Science Division, Lawrence Berkeley National Laboratory, Berkeley, CA 94720, USA; email: hlcrawford@lbl.gov}
\affil{$^2$Department of Physics, Florida State University, Tallahassee, FL 32306, USA; email: kfossez@fsu.edu}
\affil{$^3$Physics Division, Argonne National Laboratory, Lemont, Illinois 60439, USA}
\affil{$^4$Department of Physics, North Carolina State University, Raleigh, NC 27695, USA; email: skoenig@ncsu.edu}
\affil{$^5$Department of Physics and Astronomy, Michigan State University, East Lansing, MI 48823, USA}
\affil{$^6$Facility for Rare Isotope Beams, East Lansing, MI 48823, USA; email: spyrou@frib.msu.edu}}

\begin{abstract}
The field of nuclear science has considerably advanced since its beginning just over a century ago.
Today, the science of rare isotopes is on the cusp of a new era with theoretical and computing advances complementing experimental capabilities at new facilities internationally.
In this article we present a vision for the science of rare isotope beams (RIBs).
We do not attempt to cover the full breadth of the field, but rather provide a perspective and address a selection of topics that reflect our own interests and expertise.
We focus in particular on systems near the drip lines, where one often finds nuclei that are referred to as ``exotic,'' and where the role of the ``nuclear continuum'' is only just starting to be explored.
An important aspect of this article is the attempt to highlight the crucial connections between nuclear structure and nuclear reactions required to fully interpret and leverage the rich data to be collected in the next years at RIB facilities.
Further, we connect the efforts in structure and reactions to key questions of nuclear astrophysics.
\end{abstract}

\maketitle

\tableofcontents

\section{INTRODUCTION}
\label{sec:Intro}

It has been 127 years since Becquerel first observed what we now know to be radioactivity, and 74 years since the parallel development of the nuclear shell model by Jensen and Goeppert Mayer~\cite{Goeppert49_421,Jensen04_233} in 1949.  Just two years after this milestone, in 1951, short-lived isotopes of krypton (Kr) were produced and separated~\cite{CERN-Kr}, demonstrating a precursor to the Isotope-Separation On-Line (ISOL) technique for producing rare isotopes. The continued development of heavy-ion accelerators opened the study of neutron-deficient isotopes populated in fusion-evaporation measurements in the 1960s and then in the 1990s the field of nuclear physics moved toward neutron-rich nuclei with new ISOL and fragmentation facilities across the globe beginning operations~\cite{Thoennessen04_1165}. 

In this relatively short period of time, nuclear physics and the science of rare isotopes has developed at an impressive pace.  The atomic nucleus is now not only viewed as a unique laboratory for understanding the fundamental nature and origin of matter, but also as a window into aspects of a broader class of quantum systems. 

In parallel, the field of astrophysics has raced forward.  The stage was set in 1957 with the seminal work of Burbidge, Burbidge, Fowler, and Hoyle exploring the idea of chemical synthesis in stars~\cite{B2FH}. Sixty years later, the first observation of gravitational waves from the merging of two neutron stars and the associated $\gamma$-ray burst, demonstrated the modern capability of multi-messenger astronomy, providing unparalleled insight into the processes relevant at astrophysical sites.

Importantly, the field of nuclear physics does not show any signs of a slowing rate of progress.  Experimentally, facilities available for measurements of rare isotopes continue to develop and extend their reach. The Rare Isotope Beam Factory (RIBF) in Japan, SPIRAL2 in France, the Facility for Antiproton and Ions Research (FAIR) in Germany, as well as the Advanced Rare Isotope Laboratory (ARIEL) at TRIUMF (Canada) are all either running or scheduled to begin operation soon, and they will all expand both the reach of nuclei studied in experiments, as well as the techniques for doing so.  In the U.S., the Facility for Rare Isotope Beams (FRIB) began operation in early 2022 and will ramp up beam power over the next several years to ultimately provide access to thousands of previously unstudied isotopes.  Coupled with cutting-edge detector systems, the experimental discovery potential of the next decade in the areas of nuclear structure, nuclear reactions, and nuclear astrophysics is unparalleled in the history of the field.

On the theoretical front, the progress and momentum is equally exciting. \textit{Ab initio} computations of \isotope[208]{Pb} were published in 2022~\cite{Hu2022:208Pb}, while in 2015 the state of the art for this type of calculation was limited to closed shells below $N,Z$ = 30.  In parallel, \textit{ab-initio} reaction theory also made impressive progress, most recently reviewed in Ref.~\cite{Navratil:2022lvq}.

\begin{figure}[htbp]
\includegraphics[width=3in]{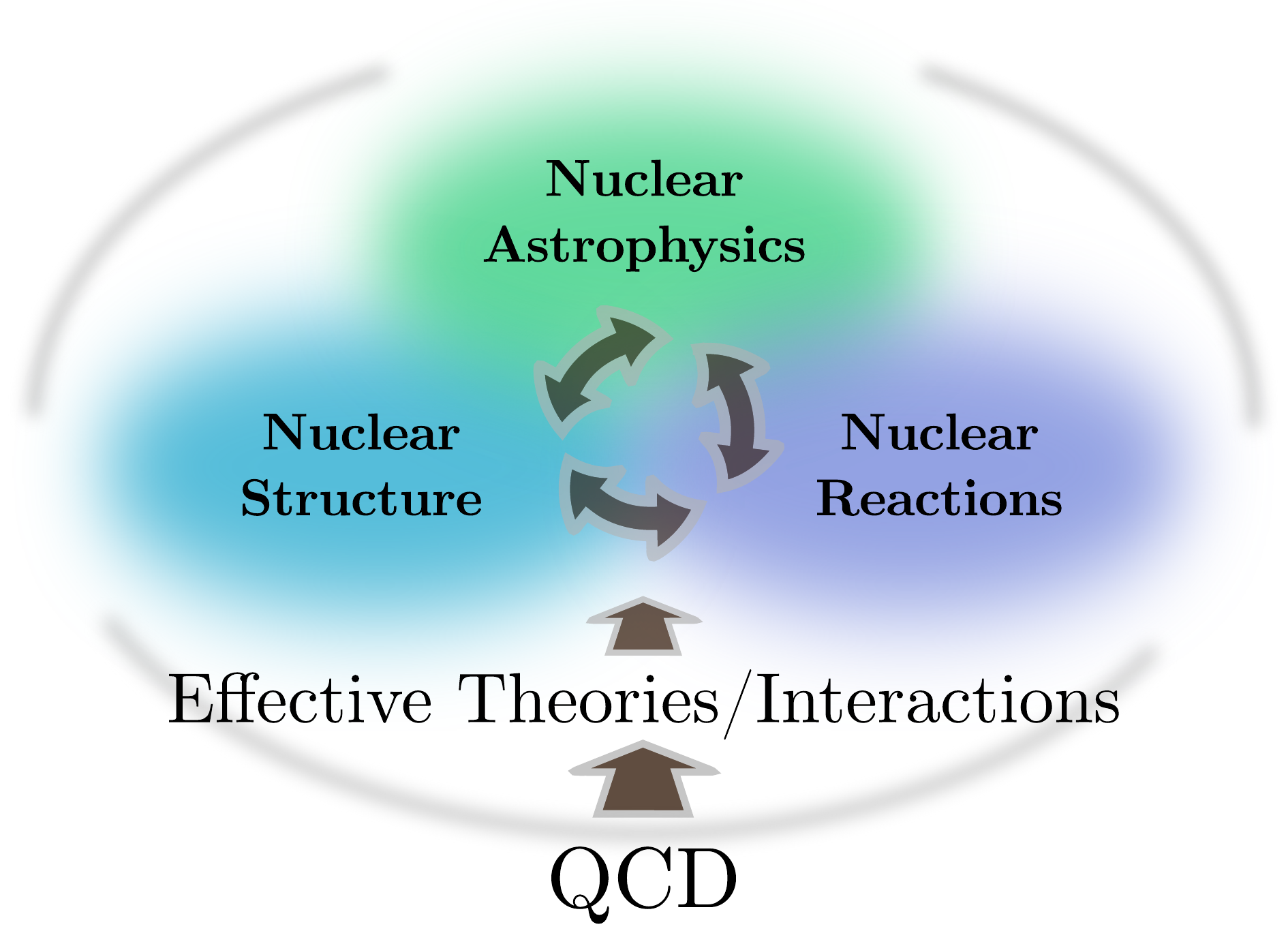}
\caption{A schematic overview of this article, illustrating the intersection of nuclear structure, nuclear reactions, and nuclear astrophysics, which are all underpinned by effective theories and nuclear interactions, and ultimately connected to QCD.}
\label{fig:overview}
\end{figure}

As illustrated schematically in Fig.~\ref{fig:overview}, we discuss in this article the intersection of nuclear structure, nuclear reactions, and nuclear astrophysics, all of which are ultimately connected to Quantum Chromodynamics (QCD), and more generally to the Standard Model (SM) of particle physics.  Effective field theories (EFT) and approaches inspired by EFT concepts provide the bridge from these underlying fundamental theories to the rich range of emergent phenomena observed in exotic nuclei.

Adopting a forward-looking approach, we present in the following sections our vision for how this path may be followed and what the next period of research may add to the picture.  This work is not intended to cover the full breadth and history of the field, but rather reflects the specific topics and expertise of the authors.  We focus, in particular, on systems near the drip lines, where the impact of near-threshold effects on nuclear properties remains to be explored in detail.  We also highlight the crucial connections between nuclear structure and reactions required to fully interpret and leverage the rich data that will be collected in the coming years.  Finally, we link the efforts in nuclear structure and reactions to key questions of nuclear astrophysics, and highlight areas in which we see the potential for advancements.

\section{EMERGENT PHENOMENA IN EXOTIC NUCLEI AND EXPLORATION OF THE DRIP LINES}
\label{sec:emergent-phenomena}

Emergent phenomena are recurrent in nuclear physics and play a prominent role in what are commonly referred to as ``exotic nuclei'' (see Sec.~\ref{sec:exoticExperiment}).  They notably include the evolution of shell structure with isospin and excitation energy, the emergence of collective degrees of freedom, but also exotic structures in states near the edges of nuclear stability, the so-called ``drip lines.''

The exploration of the drip lines is a major science driver in low-energy nuclear physics, and, to borrow from Ref.~\cite{Thoennessen04_1165}, stems from the deceptively simple question: ``Which combinations of neutrons and protons can form a nucleus?'' -- where a ``nucleus'' in this context means a system bound with respect to nucleon/cluster emission.
From a theoretical point of view, this question, at a trivial level, is asking where the one-neutron and one-proton separation energies, $S_n$ and $S_p$ respectively, cross zero.
Given that in stable nuclei the typical binding energy per nucleon $E/A \approx 8.0$ MeV is orders of magnitude below the nucleon mass $m \approx 1.0 \text{GeV}/c^2$, one can easily see how finding the $S_{n,p} = 0$ ``drip lines'' is inherently a low-energy nuclear physics problem.
At a deeper level, however, this question is really asking how the complexity of QCD at low energy and the generic properties of fermionic many-body open quantum systems together shape the limits of the nuclear landscape.
The current paradigm in low-energy nuclear theory, which we discuss more in Sec.~\ref{sec:EFT}, is to answer this question by constructing effective nuclear forces derived from QCD in the EFT framework, and then to use these forces to solve the \textit{ab initio} quantum many-body problem in a unified picture of nuclear structure and reactions.
However, despite the impressive developments of the past few decades, reminiscent of the Hydra in ancient Greek mythology, every time some progress is made in the exploration of the drip lines, a multitude of new challenges emerge due to the uncanny complexity of the atomic nucleus.

Experimentally, the race for determining the limits of the nuclear chart has always been a driver for new detector systems and new facilities~\cite{Thoennessen14_1552}.  As a case in point, FRIB is opening a new era for low-energy nuclear physics, together with other current and upcoming facilities, as mentioned in Sec.~\ref{sec:Intro}.
By itself, FRIB and the planned FRIB 400~MeV/u energy upgrade will triple or even quadruple the number of isotopic chains for which the neutron drip line is experimentally accessible, thereby extending our knowledge of this line from $Z=10$ (Ne)~\cite{Ahn:2022ribf} up to $Z$=30-60~\cite{FRIB400}.
It will also provide increased intensities at, or beyond, the proton drip line.
The discovery of new exotic isotopes in extreme $N/Z$ conditions is expected to reveal new phenomena that challenge current theoretical and experimental paradigms. 

Already, new tools are being developed to observe and describe exotic structures in nuclei, but before going over current efforts related to the exploration of the drip lines and speculating over what the future may hold, for context, we will first provide a brief account of what we have learned in the past few decades by moving away from the valley of stability.

\subsection{Historical perspective}
\label{sec:History}

Following the landmark discovery of the atomic nucleus by Rutherford in 1911~\cite{Rutherford11_783}, it took about half a century and more than a dozen Nobel prizes to set the foundations of nuclear physics and develop the necessary experimental tools for the field. Around 1950, the establishment of the shell structure of the nucleus by Goeppert-Mayer, Haxel, Jensen, and Suess~\cite{Goeppert49_421,Haxel49_420} and the development of collective models by Rainwater, Bohr, and Mottelson~\cite{Rainwater50_1268,Bohr52_1252,Bohr53_1251} defined two seemingly contradictory views of the nucleus with far-reaching consequences~\cite{Heyde13_399}.
The reconciliation came soon after, when Elliott and Flowers described the emergence of deformation in \isotope[19]{F} from single-particle degrees of freedom in the spherical shell model~\cite{Elliott55_2863}, leading to the formulation of Elliott's $SU(3)$ model~\cite{Elliott58_2469,Elliott58_2470} that describes deformed solutions in the intrinsic frame via collective couplings without any symmetry breaking in the laboratory frame. 

Even though it lacked a fully microscopic foundation, the nuclear shell model became a tool of choice due to its practical success and its intuitive appeal. 
However, to quote Bethe~\cite{Bethe56_2440}, \emph{``Nearly everybody in nuclear physics has marvelled at the success of the shell model.''}
Indeed, it was unclear how low-lying nuclear states could be described in a picture in which nucleons evolve on well defined orbits while having a Fermi momentum around 350 MeV and a highly repulsive interaction at very short distances. 
The idea that Pauli's principle was responsible for this situation had already been suggested, but it was Brueckner who first showed that, in infinite uniform nuclear matter, large cancellation effects are indeed at play and effectively result in a renormalization of nuclear forces~\cite{Brueckner55_2503}. The proof was later extended and simplified to finite nuclei by Bethe~\cite{Bethe56_2440} and Goldstone~\cite{Goldstone57_2916}. 

To complete the foundation of low-energy nuclear physics, a theory of the underlying nuclear forces was necessary. Following the groundbreaking work of Yukawa in 1934~\cite{Yukawa55_2906}, meson-exchange theories were proposed with various degrees of success, until the discovery of QCD, which happened to be non-perturbative at low energy, rendered the problem seemingly unsolvable. More detailed historical accounts of these developments can be found in Refs.~\cite{Epelbaum09_866,Machleidt:2017vls}.

To add insult to injury, connecting these unknown residual nuclear forces of QCD to properties of nuclei requires solving the fermionic quantum many-body problem, which in itself remained too difficult to handle beyond a few nucleons for decades. 
This ``crisis'' of nuclear physics and its ultimate resolution is what led to some of the current efforts mentioned in this review. 
While going through all the developments of the 1960-1990 period is far beyond the scope of this review, we highlight in the following selected examples that have particularly strong connections to current efforts.

\textbf{Shell evolution.}  As new experiments gave access to nuclei away from the valley of stability in the late 1980s and 1990s, changes in the traditional shell structure started to become apparent. The empirical shell model~\cite{Brown01_995,Caurier05_424}, based on single-particle energies and two-body matrix elements within a given model space and optimized on many-body data, transitioned to increasingly more sophisticated and precise models culminating with the USD family of interactions~\cite{Brown06_1828} and similar models for the $fp$ shells~\cite{Otsuka01_2384,Caurier05_424}. 
This approach not only provided invaluable support for experiments at a critical time, but also demonstrated that the renormalization of nuclear forces in nuclear matter, uncovered by Brueckner, can be effectively achieved within the shell-model framework well beyond expectations. 
From there, many theoretical developments followed to derive shell-model Hamiltonians directly from \textit{ab initio} forces~\cite{Stroberg19_2314}, \textit{i.e}, adjusted only on few-body data, which opened the door to direct comparisons between shell model and \textit{ab initio} calculations, but also to consistent calculations of observables within the shell model framework~\cite{Stroberg19_2314}. 

These shell model developments made it possible to understand the emergence of Islands of Inversion (IOIs) on the nuclear chart, where nuclear structure deviates from the standard shell model predictions due to the evolution of shell structure, largely driven by tensor forces~\cite{Otsuka20_2383}.  See Sec.~\ref{sec:exoticExperiment} for details.

\textbf{Nuclear halos.}  Halo structures, discovered in 1985, are among the most emblematic exotic phenomena uncovered by the exploration of the drip lines~\cite{Tanihata13_549}. 
Their discovery triggered a wave of new experimental techniques and programs at facilities such as RIKEN, GANIL, and NSCL, to not only search for new halo states but also to extend the limits of their early definition. 
The realization that weak binding could produce extended structures in nuclei~\cite{Hansen87_231}, characterized by the emergence of new effective scales and associated degrees of freedom, led to the introduction of scaling laws~\cite{Jensen04_233} and the general concept of universality~\cite{Canham08_2450,Frederico12_372}, as well as the development of effective field theories for halo systems~\cite{Bertulani:2002sz,Hammer:2017tjm,Hammer:2019poc}, which we discuss further in Sec.~\ref{sec:HaloEFT}.

\textbf{Nuclei as open quantum systems.}  More generally, going away from the valley of stability revealed that other near-threshold effects besides halos were important to understand exotic nuclei, such as 
near-threshold clustering~\cite{Michel10_4,Okolowicz12_998,Kravvaris19_2326}, or low-$\ell$ shell evolution~\cite{Hamamoto04_2200,Hoffman14_1868}, and that these phenomena could be understood as generic phenomena in the open quantum system (OQS) framework describing quantum systems coupled to an environment of decay channels and scattering states. 
The description of nuclei as OQSs~\cite{Moiseyev98_92,Okolowicz03_21,Civitarese04_34,Michel21_b260} emphasizes the role of continuum couplings in the dynamic of, for instance, exotic decay modes~\cite{Pfutzner12_1169,Thoennessen13_1776,Pfutzner23_2859}, overlapping resonances and superraddiance~\cite{Auerbach11_879,Eleuch14_860}, or trapped resonances. 
In this picture, exotic nuclei, unlike stable nuclei which are isolated from each other, are coupled through capture and decay, and must therefore be described in a unified theory of nuclear structure and reactions~\cite{Michel09_2,Volya05_470,Navratil16_1956}. 
Ironically, while this review is about rare isotope physics, it must be emphasized that excited states in stable nuclei are governed by the same near-threshold physics as exotic nuclei, modulo the extreme $N/Z$ ratio, making stable-beam facilities such as the ATLAS and ARUNA laboratories ideal places to perform precise studies to thoroughly test theoretical concepts and methods before being applied on drip-line systems.

\textbf{Nuclei as multi-scale objects.} The seemingly hopeless problem of deriving nuclear forces compatible with QCD was formally solved by Weinberg and others
with the formulation of a low-energy EFT of nucleon-nucleon interactions
based on the approximate chiral symmetry of QCD~\cite{Weinberg:1990rz,Rho:1990cf,Weinberg:1991um,Ordonez:1992xp,Weinberg:1992yk,vanKolck:1993ee}. This led to the development of so-called ``chiral potentials''~\cite{Epelbaum09_866,Machleidt11_414} that started rivaling with phenomenological high-precision potentials such as the AV18 interaction~\cite{Wiringa95_1072}.
As discussed in more detail in Sec.~\ref{sec:EFT}, fundamentally, the EFT program is akin to improving order by order the energy or momentum resolution at which one describes the system of interest. 
The parallel with existing renormalization techniques~\cite{Suzuki80_928}, and more generally with the renormalization group (RG) philosophy, was promptly noted and led to the formulation of the similarity renormalization group (SRG) method~\cite{Bogner10_961}. In the SRG method, the momentum cutoff takes a meaning equivalent to the breakdown scale in EFT. The high-momentum contributions of nuclear forces that are beyond the cutoff are absorbed into the low-momentum part while preserving few-body observables. Applied, in particular, to phenomenological potentials, this SRG ``evolution'' effectively ``softens'' the interaction by removing the ``hard'' repulsive core. At a practical level, SRG made computationally costly configuration-interaction (CI) calculations such as the no-core shell model converge faster in smaller model spaces, which in turn opened new possibilities to test nuclear forces using exact methods. 

In parallel to these developments, the re-introduction of the coupled clusters approach in nuclear physics~\cite{Hagen14_939} made it possible to express the Hamiltonian with respect to a mean-field reference state on top of which additional many-body correlations are built within the Hamiltonian via successive excitations. 
In other words, the many-body problem is solved in the Heisenberg picture instead of the Schr\"odinger picture. 
The paradigm shift lies in the ability to truncate at the level of correlations instead of configurations, as in usual CI approaches. 
Indeed, provided that a good mean-field reference state can be found, only a modest computational effort is required to capture most of the missing correlations, making the coupled clusters approach scale polynomially rather than factorially with the nuclear mass.
As a consequence, the reach of \textit{ab initio} calculations exploded! 
Surprisingly, the extension of SRG to many-body systems, giving the in-medium SRG (IMSRG) method~\cite{Hergert16_1673}, appeared to be practically equivalent to the coupled clusters approach. 
Detailed studies of the IMSRG method demonstrated how the renormalization shifts the strength of, for instance, three- and two-body operators entering the Hamiltonian toward one- and zero-body operators, \textit{i.e.}, how many-body correlations are effectively absorbed into the mean-field. 
Following these findings, several \textit{ab initio} methods were developed that could exploit truncations in many-body correlations and therefore contribute to the study of exotic nuclei. 
The EFT/RG paradigm had, and continues to have, a profound impact on how we understand nuclei as multi-scale systems, and opened the possibility to test nuclear forces derived from QCD on exotic nuclei.

To summarize, the study of rare isotopes away from stability has already challenged the original paradigms, set at the foundation of nuclear physics, through the discoveries of, for instance, the IOIs and their origin in the shell evolution, halo structures, or new forms of radioactivity, but also, and perhaps more importantly, by forcing us to look at nuclei in new ways. The description of nuclei as multi-scale objects and/or as open quantum systems is still developing and will likely be impacted by the coming online of new facilities including FRIB that will push the drip lines well beyond our current knowledge and greatly extend our knowledge in little-known regions.
It is the purpose of this review to provide a vision for the science of rare isotope beams in light of past and current developments. 

\subsection{Breadth of structural phenomena in exotic nuclei}
\label{sec:exoticExperiment}

Exotic nuclei are usually understood as nuclei that are not commonly found in Nature, \textit{i.e.} that are not stable. 
However, away from the valley of stability, one can roughly define three different regions from a structure point of view. 

\textbf{Deeply bound region.}  The first region includes systems that are predominantly unstable with respect to $\beta$ decay but remain well bound with respect to nucleon emission. 
Usually, nuclei in this region support bound excited states and can present an isospin imbalance, defined as $N/Z$ and $Z/N$ on the neutron-rich and proton-rich sides, respectively, ranging between about 1.0 and 3.0. In this vast region, early measurements of masses, ground-state spins, and magnetic moments~\cite{Thibault.1975,Huber.1978} in the Na ($Z=11$) isotopes near $N=20$, as well as low-lying excited state energies in \isotope[32]{Mg}~\cite{Detraz.1979} revealed the first break in the standard shell model paradigm, which was interpreted as evidence of deformation in what is now known as the $N=20$ IOI~\cite{Wildenthal80_1522,Poves87_1500,Warburton:1990}. 
With the development of progressively more powerful radioactive ion beam facilities, new IOIs were identified at $N=8$, 14, 20, 28, 40 and 50~\cite{Brown2010Physics,Nowacki2016:N50}. 
The observed disappearance of the shell model ``magic numbers'' away from stability, associated with expected shell closures and increased stability in this approach, was accompanied with the observation of new unexpected sub-shell closures, such as the $N=32$ and 34 sub-shell closures in \isotope[52]{Ca}~\cite{Huck1985.52Ca,Gade2006.52Ca} and \isotope[54]{Ca}~\cite{Steppenbeck2013}.

With new RIB facilities coming online, the number of isotopes experimentally accessible, currently estimated at about 3000, is expected to roughly double. 
Masses and level excitation energies, determined in decay or reaction studies, will most likely constitute the first information available to see new changes in the structure of nuclei at large isospin imbalances. 
Later, detailed spectroscopic studies will provide additional information on the discrete spectrum of exotic nuclei, which can be used to reveal phenomena such as shape coexistence~\cite{Garrett2022:shapecoex} or octupole deformation~\cite{Ahmad1993:Octupole}, which were once thought to be rare but might in fact be quite common~\cite{Rowe10_b210}.
At an even more refined level, transition matrix elements, which can be accessed experimentally by excited state lifetime or Coulomb excitation measurements, can provide detailed information about the structure of nuclei, which can then be compared with \textit{ab initio} calculations as was done, for instance, in Refs.~\cite{Ciemala2020:20O16C, Heil:202021C} for carbon and oxygen isotopes. 
One notes that, in heavy nuclei, the experimental challenge posed by more complex level schemes will be met by the high resolution of next-generation gamma-ray spectrometers such as AGATA~\cite{AGATA} and GRETA~\cite{GRETA}. 

An open question is whether the shell evolution driven by tensor forces and the mass-dependence of the mean-field will continue to be the sole drivers behind the emergence of IOIs and new shell closures, or if new effects will become apparent under extreme $N/Z$ conditions, even for systems that remain relatively well bound. 
Somewhat related to shell evolution, it remains to be seen if a dramatic enlargement of the neutron skin is to be expected in systems beyond \isotope[60]{Ca}, \isotope[78]{Ni}, or \isotope[132]{Sn}, and how these new data will constrain the nuclear equation of state and our understanding of neutron stars~\cite{Thiel19_2981,Lattimer21_2982}. 

Another topic of great interest concerns the tendency of nuclei to form $\alpha$ clusters. 
It has been shown that nuclear matter appears to be near a phase transition between a nuclear liquid and a Bose condensate of $\alpha$ clusters~\cite{Elhatisari16_2311}, but also that the Wigner-$SU(4)$ symmetric part of nuclear forces, that are mostly responsible for the binding of $\alpha$ clusters, largely controls nuclear binding in medium-mass and heavy nuclei~\cite{Lu:2018bat}. 
It will be interesting to see how nuclear binding evolves under an increasingly large excess of neutrons, and whether or not this evolution becomes closer to a purely Wigner-$SU(4)$ picture, potentially making $\alpha$-cluster correlations within the nucleus more important. 
In either case, better constraining nuclear forces on emergent phenomena such as deformation and clustering in light nuclei might prove critical to understand the dynamics of heavier systems. 
Finally, one can also speculate about the possibility that past a certain neutron excess, deformation could develop not due to traditional factors related to the proton-neutron interaction~\cite{Federman79_2361,Dobaczewski88_1584,Pittel93_2367,Nazarewicz94_1250}, but due to somewhat delocalized pairs of neutrons at the surface, similarly to what might be happening in \isotope[8]{He}~\cite{Holl21_2507} or \isotope[40]{Mg}~\cite{Crawford:2019Mg}. 

\textbf{Weakly bound region.}  The second region is reached when the one-neutron or one-proton separation energy falls below $\approx 1.0$ MeV and is thus about one order of magnitude smaller than the average binding energy per nucleon in nuclear matter $B(A,Z)/A \approx 8.5$ MeV, and ends in a fuzzy manner when $\Gamma/(2 S_{n/p}) \approx 1.0$ in the ground state \cite{Michel21_b260}, \textit{i.e.} the ground state cannot be reasonably described as a narrow resonance anymore, at least in a theoretical sense. 
This is the region around the drip lines characterized by near-threshold physics. 
While near-threshold physics is not limited to the drip lines and is in fact relevant in many excited states of well bound nuclei, exotic nuclei in this region present unique opportunities to test nuclear forces in extreme $N/Z$ conditions due to the sensitivity of near-threshold phenomena. 

For example, as mentioned earlier, below the particle emission threshold weak binding can lead to the formation of halo structures, providing that the centrifugal barrier is either nonexistent or low. 
The spatial extension of halo states strongly depends on the binding energy of halo nucleon(s) and, in addition, in the case of halos involving more than one nucleon, angular correlations between halo nucleons can change dramatically depending on the partial waves involved and the presence or not of Coulomb forces. 
In the simplest case of two-neutron halos, such correlations manifest themselves by forming so-called ``cigar'' \textit{vs.} dineutron configurations~\cite{Bertsch91_2181,Hagino05_2666}. 
The presence of ground-state halos in this region is clearly manifest in Fig.~\ref{fig:driplines}. 

Above the threshold, in addition to virtual couplings to higher energy scattering states, couplings to lower energy states open, leading to the formation of resonances and the phenomenon of particle decay. 
Resonances, characterized by an energy position and a width, can be isolated from each other in energy like bound states, but also overlap and strongly couple through the continuum of scattering states, leading to the phenomenon of superradiance~\cite{Auerbach11_879,Eleuch14_860}. 
Continuum couplings also appear between different partitions of a nucleus through decay channels and lead to few-body decay. 
For example, depending on the energy pattern of a given isotopic chain, new exotic forms of radioactivity can become dominant such as two-neutron or two-proton decay~\cite{Pfutzner23_2859}. 
To date, the most exotic form of few-body decay observed is the five-proton decay of \isotope[9]{N}~\cite{Charity23_2960}. 

Somewhat related to this phenomenon is the tendency of nuclei to form resonances near decay channels. 
Naturally, the wave function of such near-threshold state ``aligns'' with the partition of the nearby decay channel, leading to the phenomenon of near-threshold clustering~\cite{Michel10_4,Okolowicz12_998,Kravvaris19_2326}. 
Why nuclei form such states near decay threshold remains unclear, but the phenomenon can have consequences on low-energy capture cross sections relevant for nuclear astrophysics, and is also at the origin of so-called ``trapped'' resonances, which are resonances well above a certain decay channel that present an abnormally small decay width in this channel because their wave function is aligned with a second closer decay channel associated with a different partition of the system. 

In the era of RIB facilities in which the drip lines will be pushed well beyond our current knowledge, one could expect the near-threshold region of the nuclear chart to broaden on the neutron side as the mass increases. 
Indeed, when approaching the drip line the binding energy tends to flatten as the wave function reorganizes to accommodate continuum couplings~\cite{Dobaczewski94}. 
If this is the case, exotic decay modes involving many nucleons could become common in the medium-mass region and beyond, presenting ideal conditions for the emergence of universal phenomena. 

\textbf{Unbound region.}  The last region is concerned with broad resonances and ends with the limits of nuclear existence, \textit{i.e.} when the lifetime of the ground state is comparable to the times it takes a nucleon to complete an orbit inside the nucleus or about $10^{-22}$s~\cite{Goldanskii66_1159,Thoennessen04_1165,Fossez16_1335}. 
In this region, strong continuum couplings dominate and time-dependent approaches formulated in the language of reaction theory are often more appropriate.  For example, it is known that at very early and long times, decay is non-exponential~\cite{Moshinsky52_1650,Khalfin58_1360}, and broad resonances, which can be understood as states in which the non-resonant part is important if not dominant, have the potential to reveal non-exponential decay features by magnifying the interference between the resonant and non-resonant parts of the wave function at long decay times~\cite{Peshkin14_1051,Wang23_2971}. 
Broad resonances can appear trivially due to their distance to a threshold, but also for more complex reasons. 
In the superraddiance phenomenon mentioned previously, a very broad (superradiant) state can be formed when the widths of many states with same spin-parity is ``collectivized'', \textit{i.e.}, concentrated in one state while all the other states become narrow. 

A common reason for the presence of broad resonances is also the opening of several decay channels.
In neutron-rich systems, the competition between multiple decay modes often leads to sequential decay involving broad resonances. 
Such many-neutron resonances could provide precious information about the poorly known neutron-neutron interaction~\cite{Hammer14_1203,Revel18_2022,Kubota20_2411,Gobel21_2607,Yamagami22_2677,Corsi23_2839}. 

Going into the extreme, experimental attempts to form a four-neutron system or tetraneutron by leveraging the four-neutron halo structure of \isotope[8]{He} led to the observation of a low-energy peak in the cross section~\cite{Kisamori16_1463,Duer22_2494}, prompting speculations about the existence of a four-neutron resonance~\cite{Hiyama16_1624,Shirokov16_1791,Fossez17_1916,Gandolfi17_2143,Li19_2634}. 
Later investigations demonstrated that the peak was only due to the residual interaction between the four neutrons in the presence of the $\alpha$, but that no proper tetraneutron state could actually form~\cite{Deltuva18_2079,Higgins20_2495,Lazauskas23_2744}. 
A more promising avenue to test neutron-neutron forces in extreme conditions is the hydrogen chain, where the single proton provides the necessary binding to form at least \isotope[5]{H} resonance states according to both theory~\cite{Lazauskas18_2032,Lazauskas19_2363} and experiment~\cite{Wuosmaa17_1940}, and perhaps even \isotope[7]{H} states~\cite{Caamano07_991,Caamano08_1938,Bezbakh20_2381,Muzalevskii21_2546}. 

\subsection{Paradigm shift at the drip lines}
\label{sec:driplineExperiment}

\begin{figure}[h]
\includegraphics[width=5in]{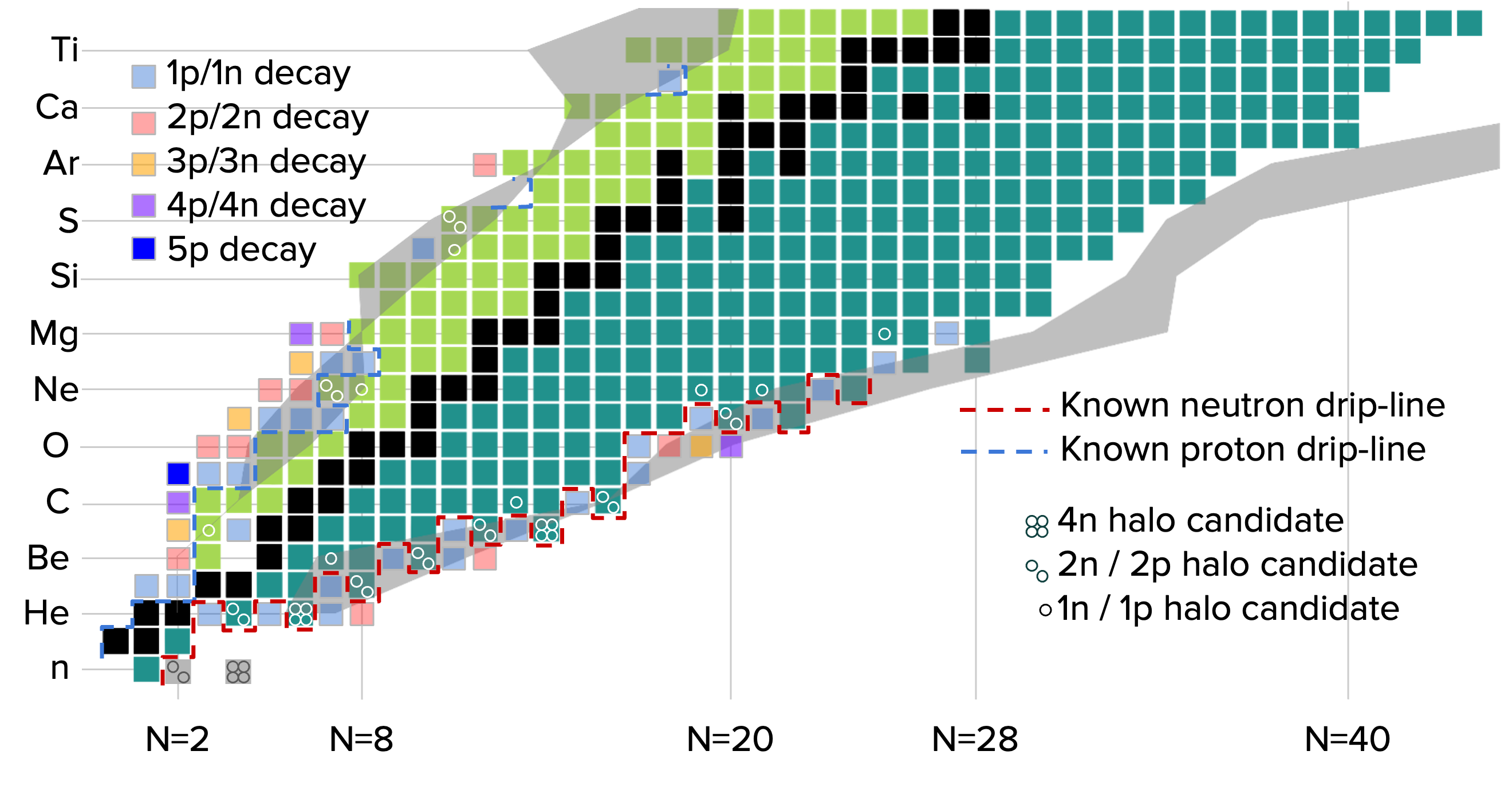}
\caption{Current knowledge of the drip lines up to $Z$=22 (Ti).  The neutron drip line is confirmed up to $Z$=10, while the proton drip line is stringently defined as far as $Z$=13 and locally at heavier masses. }

\label{fig:driplines}
\end{figure}

Our current experimental knowledge of the drip lines is surprisingly limited.  It took almost 50 years to extend the neutron drip line up to oxygen ($Z=8$) isotopes, and 20 more years to push it up to neon ($Z=10$) isotopes~\cite{Ahn19_2387}.  The proton drip line is also difficult to determine experimentally due to the Coulomb barrier that can extend the lifetime of proton-unbound nuclei by many orders of magnitude~\cite{Thoennessen04_1165}. 
It has been crossed up to $Z=83$, but only established stringently up to $Z=13$~\cite{AME2020}.
For context, the largest isotopes ever created, \isotope[294]{Og}, has $Z=118$. 
A summary of our knowledge of the drip lines below $Z=25$ is shown in Fig.~\ref{fig:driplines}. 

As mentioned in Sec.~\ref{sec:exoticExperiment}, the region around the drip line presents many new phenomena related to near-threshold physics, which in itself is already a paradigm shift since it leads to the idea of the unification of nuclear structure and reactions for the description of nuclei as open quantum systems. 
However, this is probably not the end of the story and, in fact, recent experimental results in systems such as \isotope[8]{He}~\cite{Holl21_2507}, \isotope[28,29]{F}~\cite{Revel20_2372,Bagchi20_2356}, or \isotope[40]{Mg}~\cite{Crawford:2019Mg}, are pointing toward the idea of \emph{interplay of continuum couplings and emergent phenomena.} 

Emergent phenomena such as the formation of a self-consistent shell structure, pairing, deformation, clustering, and collective motion are well known, even though not fully understood. 
Their very existence is in large part what makes the atomic nucleus a complex system. 
In the absence of a clear separation of scales between two or more such phenomena, subtle interplay can emerge like, for instance, when single-particle and collective dynamics compete in a particle-plus-rotor type of system, or when pairing and deformation compete to lower the energy of a nucleus. 
The accurate description of such interplay from first principles remains a challenge~\cite{Launey16_2403}, largely due to the need to fully capture both static (mean-field) and dynamic (particle-hole) correlations in the many-body description. 

Near the drip lines, the possibility of particle decay via continuum couplings lead to new types of interplay by changing the priorities of the system. 
Indeed, above the particle-emission threshold, quantum systems obey the generalized variational principle which lowers both the energy and the decay width of the system~\cite{Moiseyev98_92}. 
A given emergent phenomenon will not be promoted solely because it lowers the energy of the system. 
Depending on the situation, it could be preferable to increase the energy in order to reduce the width by a larger factor. 
For example, in a particle-plus-rotor picture, these phenomena can generate narrow neutron resonances in \isotope[11]{Be} by forcing the valence neutron to occupy high-$\ell$ states when \isotope[10]{Be} rotates faster~\cite{Garrido13_1171,Fossez16_1335}. 

Below the threshold, while the dynamics of weakly bound systems is, by definition, not directly affected by decay, the relatively small energy cost of continuum couplings, in this case, effectively forces the wave function of weakly bound systems to spatially delocalize. 
One way to understand intuitively this phenomenon is to picture a quantum system whose binding energy is reduced until it decays by particle emission. 
The distribution of the particles in the system must remain continuous at the threshold. 
Thus one could say that, right below the threshold, the system must ``prepare'' to decay. 
Halo states are a prime example of this phenomenon. 

An illustration of the interplay between continuum couplings and emergent phenomena can be found in the ground state of \isotope[8]{He}. 
The four halo neutrons, each bound by about 0.7 MeV to a tightly bound core of \isotope[4]{He}, occupy low-$\ell$ states and are well delocalized in space.  Yet, the presence of significant deformation was recently reported~\cite{Holl21_2507}. 
It appears that the deformation might be related to the emergence of dineutron correlations leading to delocalized pairs of neutrons orbiting around \isotope[4]{He}~\cite{Yamaguchi23_2970}. 
How the halo dynamics, dineutron correlations, and deformation compete remains to be fully elucidated. 

A similar situation is suspected in the ground state of \isotope[40]{Mg} which might present a two-neutron halo structure~\cite{Crawford:2019Mg} dominated by $p$-waves ($\ell=1$), with the complication that \isotope[38]{Mg} and surrounding nuclei are known to be well deformed~\cite{Doornenbal:2013Mg}. 
One notes that a one-neutron halo structure is suspected in \isotope[37]{Mg}, and that \isotope[39]{Mg} is barely unbound~\cite{Fossez16_1793}. 
Surprisingly, \isotope[40]{Mg} is sufficiently bound to support bound excited states~\cite{Crawford:2019Mg}, making the importance (or lack thereof) of the neutron continuum on this nucleus as of yet unclear. 
The deformation of the core could, in this case, lead to an increased occupation of $f$-waves ($\ell=3$) in excited states. 

A different and yet related interplay of continuum couplings and emergent phenomena is found in neutron-rich fluorine ($Z=9$) isotopes, right on the edge of the $N=20$ IOI. 
The observed and expected presence of negative parity states in \isotope[28]{F}~\cite{Revel20_2372}, \textit{i.e.}, at $N=19$ and thus before the IOI, and of a halo structure in \isotope[29]{F}~\cite{Bagchi20_2356}, suggested a modification of our understanding of the IOI. 
It was proposed that already in \isotope[28]{F}, which is unbound, continuum couplings promote the occupation of $p_{3/2}$ waves, which in turn promotes couplings with $f_{7/2}$ waves leading to quadrupole deformation, and the continuum-induced deformation continues to develop in \isotope[29]{F} and heavier isotopes~\cite{Luo21_2394,Fossez22_2540}. 
While further experimental investigations are needed, fluorine isotopes provide yet another example of unique and complex interplay in exotic nuclei near the drip lines. 

Many more cases of interplay are to be expected in relation to, for instance, near-threshold clustering. 
The connection between continuum couplings and the mechanism leading to clustering in well bound systems remains to be established~\cite{Michel10_4,Okolowicz12_998,Kravvaris19_2326}. 
One can wonder if solving this problem will provide at the same time a unified understanding of halo structures and exotic decay modes, which can both be regarded as special cases of near-threshold clustering if one treats dineutron and diproton correlations as hints of fermionic pairs~\cite{Wang21_2629}.

\subsection{Towards complete measurements}
\label{sec:structureExperiment}

The previous sub-sections discussed many of the most compelling topics in the area of nuclear structure, including shell evolution and the nature of the atomic nucleus as an open quantum system.  It is important to realize that experimental studies to address these physics topics are entering a new era, including not only measurements of systems lying at, or beyond, the drip lines, but also measurements exploring the limits of nuclei closer to stability that are pushing to extremes of excitation energy, where the continuum is again a critical ingredient.  The same types of studies will also provide key insights into the structure of rare isotopes, and help to constrain the drivers of shell evolution across the nuclear chart.

At and beyond the drip lines, the intensity of primary beams at fragmentation facilities such as FRIB will rapidly extend the experimental reach toward the neutron drip line at progressively higher $Z$.  Adding to this, experimental approaches are being developed and deployed to maximize the information obtained in each experiment, with simultaneous measurement of all emitted radiations.  For example, decay spectroscopy studies, which are often among the first experiments possible for isotopes produced at the lowest yields now routinely include detectors for charged particle detection ($\beta$, $\alpha$ and proton emissions, including conversion electrons), in addition to $\gamma$-ray detection arrays and neutron detection setups.  As examples, the Isolde Decay Station at CERN-ISOLDE, the GRIFFIN facility at TRIUMF~\cite{GRIFFIN_TRIUMF} and the FRIB Decay Station Initiator (and ultimately the FRIB Decay Station) are all end-stations for truly complete decay spectroscopy.  With these extremely sensitive devices, fully correlated measurements are possible, in which decay to both bound and unbound excited states are measured with a complete accounting and detection of emitted gamma-rays and neutrons.

Beyond decay spectroscopy, advances and extensions of reaction study experimental setups will also extend the capabilities for in-beam spectroscopy and reactions experiments to probe nuclear structure.  As is the case for decay measurements, reaction studies on drip line and near drip line nuclei, including Coulomb excitation and nucleon removal/addition reactions, will be enabled and enhanced in the near future with ``complete'' experimental setups including charged-particle, neutron and gamma-ray detection.  These are already in use with R$^{3}$B~\cite{R3B} at GSI, the SAMURAI set-up at RIBF, and planned at the FRIB High Rigidity Spectrometer (HRS).  The data collected at such facilities allows simultaneous investigation of the overlap between the ground states of parent nuclei and the bound and unbound states populated in reactions of the most neutron-rich nuclei, such as the recent results for the case of proton removal from \isotope[25]{F} into bound states and resonant states in \isotope[24]{O}~\cite{Tang_25FPRL}.

Technical developments of targets will also extend the experimental reach even further from stability by optimizing the luminosity for measurements of the most exotic systems.  As an example, thick LH$_{2}$ targets such as MINOS and similar systems~\cite{MINOS,STRASSE} will maximize the total luminosity for direct reactions populating the most neutron-rich systems.
The future prospects for performing reactions \emph{on} drip line nuclei are bright. 


\section{BUILDING A RIGOROUS AND CONSISTENT PATH FROM QCD TO NUCLEAR REACTIONS}
\label{sec:Path}

As mentioned already in Sec.~\ref{sec:History}, the discovery of
QCD generated the challenge of understanding how atomic nuclei emerge out of the fundamental interactions between quarks and gluons, which at low energies are highly non-perturbative.
While lattice simulations are now able to simulate at least very light nuclei in terms of these degrees of freedom (see for example Refs.~\cite{Orginos:2015aya,Berkowitz:2015eaa,Yamazaki:2015asa}), or try to extract a baryon-baryon potential from lattice simulations~\cite{Aoki:2023qih}, covering a significant portion of the nuclear chart with calculations from first principles will, for the foreseeable future, require more pragmatic and effective approaches, which we discuss in this section.

\subsection{Effective field theories}
\label{sec:EFT}

A tremendous amount of progress in nuclear physics over the past few decades has been driven by the development and application of EFTs.
EFTs have emerged as powerful tools widely used in modern theoretical physics.
Applied to nuclear physics, where Weinberg and others pioneered the construction of an EFT of nucleons and pions in the 1990s~\cite{Weinberg:1990rz,Rho:1990cf,Weinberg:1991um,Ordonez:1992xp,Weinberg:1992yk,vanKolck:1993ee}
, EFTs come with the ambitious promise to firmly root calculations of nuclear structure and reactions
in QCD, and to make predictions with fully quantified theoretical uncertainties.
As mentioned in Sec.~\ref{sec:History}, the central idea of an EFT is the development of a formalism that is tailored for a given ``theoretical resolution'' appropriate for what one aims to describe, achieved by reducing (through effective parametrizations) unnecessary microscopic details to a bare minimum.
EFTs therefore provide a natural justification for performing nuclear physics calculations based on nucleons as degrees of freedom, even though QCD tells us that these particles themselves are built out of quarks and gluons.
The key to this is that nuclear binding energies, of the order of few MeV per nucleon, are generally small to typical QCD scales, which are of the order of 1 GeV, set, for example, by the nucleon mass.
Converting these scales to a de Broglie wave length leads to the interpretation in terms of a resolution that determines the most effective degrees of freedom.

Of course, describing nuclei in terms of nucleons is by no means a new idea, but EFTs enable doing it in a way that maintains a systematic connection to QCD as the underlying theory, used to inform the construction of the effective interactions between nucleons.
More quantitatively, the construction of an EFT rests on the separation of a typical momentum scale $Q$, characterizing the systems and processes one wishes to describe, from physics at a larger scale $\Mhi \gg Q$, which are effectively irrelevant and/or even unknown.
Information from theses scales enters in the EFT only indirectly via the so-called ``low-energy constants (LECs)'' that determine the coupling strength of interactions in the EFT.
In nuclear physics, the connection to QCD as the underlying theory is given by the fact that QCD symmetries dictate which interaction terms are present in the EFT -- and in fact it is even possible to use QCD calculations to directly determine LECs.
In a few cases, such determinations have already been achieved using lattice calculations, see for example Refs.~\cite{Savage:2016kon,Detmold:2021oro,Illa:2021clu}.
In the EFT construction, two- and many-body forces are treated in a uniform manner, and external currents (weak, electromagnetic) can be incorporated consistently within a unified scheme.
For a recent review of nuclear EFTs formulated in terms of nucleons and clusters of nucleons, see Ref.~\cite{Hammer:2019poc}.

The ``chiral EFT'' mentioned at the outset of this section constructs the strong force between nucleons in terms of increasingly complex pion-exchange diagrams, augmented by zero-range ``contact'' interactions that effectively account for effects that are not resolved explicitly (such as the exchange of higher-mass mesons).
It was subsequently realized that at sufficiently low energy one can formulate an even simpler ``pionless EFT''~\cite{Kolck99_1066,Kolck99_1979,Chen99_1053,Steele99_1701,Bedaque02_1957}, which uses contact interactions only to parameterize the strong interaction between nucleons.
Pionless EFT is in fact driven not so much by ``integrating out'' pions from chiral 
EFT, but rather by the universal physics stemming from the nucleon-nucleon S-wave scattering lengths being large compared to the typical nuclear length scale associated with pion exchange.

Setting aside that within the community of EFT practitioners there is no overall consensus regarding the \emph{implementation} of the general EFT paradigm (the details of which we do not delve into here), it is a fact that potentials describing two- and three-nucleon interactions derived from (or at least inspired by) chiral EFT are commonly used nowadays to calculate nuclear structure observables.
There exists a number of notable efforts to extend this program also to the theory of nuclear reactions, including nucleon-nucleus and nucleus-nucleus reactions, using microscopic approaches that allow for first-principles calculations; see Refs.~\cite{Deltuva:2008aa,Leidemann:2012hr,Navratil16_1956} for reviews.
However, such calculations are typically limited to relatively light nuclei, and the vast majority of nuclear reactions at present remains treated with phenomenological optical potentials (a topic we return to further below) and few-body models without clear connection to an underlying \textit{ab initio} formalism.

\begin{figure}[htbp]
\includegraphics[width=2in]{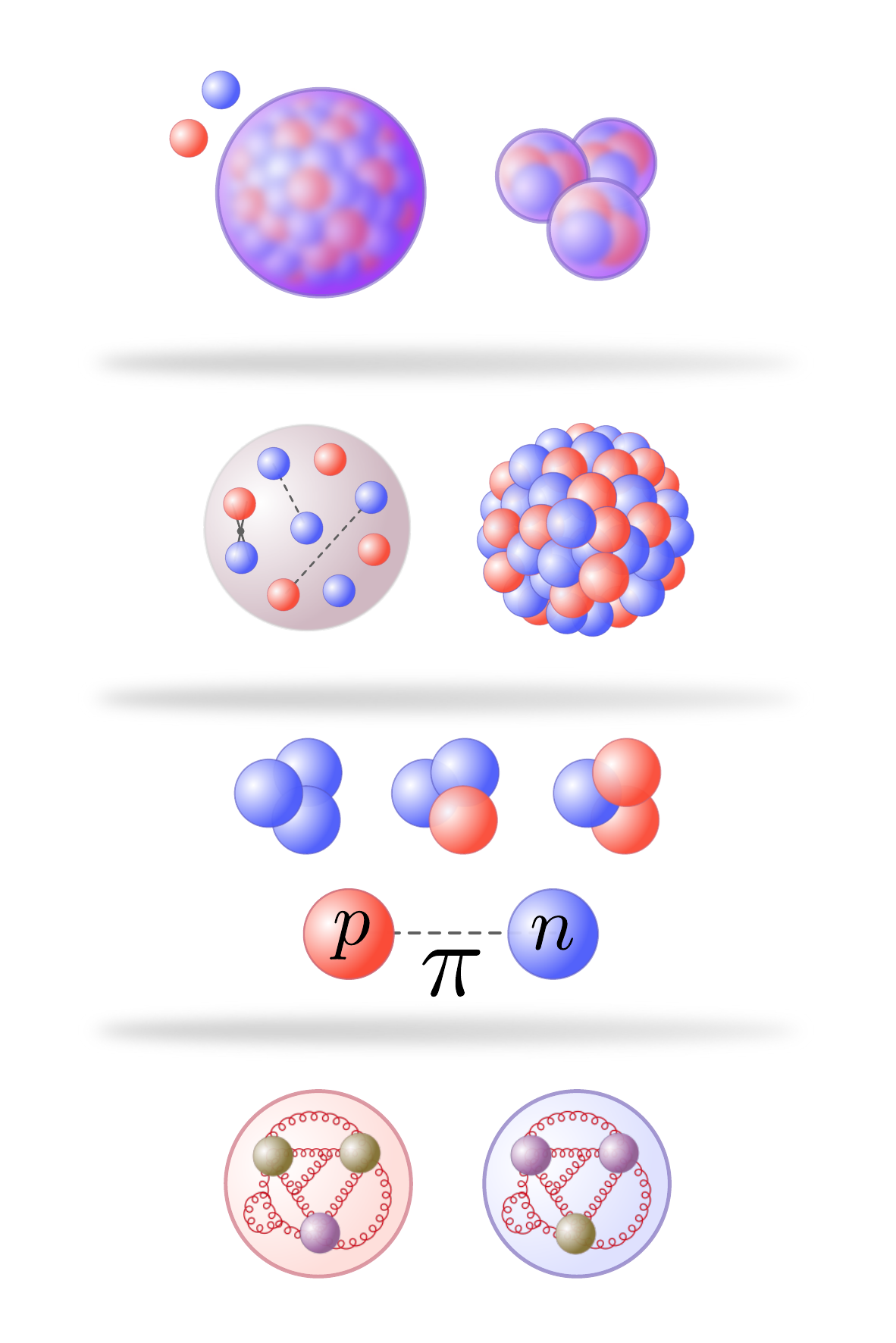}
\caption{Tower of effective degrees of freedom in nuclear theory.}
\label{fig:tower}
\end{figure}

\subsection{Halo and cluster systems}
\label{sec:HaloEFT}

For certain exotic nuclei with a pronounced halo or cluster structure, it is possible to use a framework that has become known as halo/cluster EFT~\cite{Bertulani:2002sz,Hammer:2017tjm}.
A key strength of this approach, which can be seen as a variant of pionless EFT ``lifted'' to heavier systems by including clusters of nucleons as degrees of freedom, is that it makes explicit correlations between different observables, linking, for example, low-energy capture reactions to scattering
observables. 
See for example Refs.~\cite{Zhang:2019odg,Premarathna:2019tup} for recent applications.

Calculations in halo/cluster EFT often face the challenge that the expansion parameter, given as the ratio of these energies (or some equivalent measure), can be less than an order of magnitude.
For example, for a description of the electric properties of \isotope[11]{Be}, it has been estimated to be of the order 0.4~\cite{Hammer:2011ye}.
This makes it necessary to go to higher orders in the EFT expansion to achieve reasonable precision, which, however, requires more experimental data to determine the increasing number of LECs beyond leading order (LO).

A viable path to achieving the needed progress towards precise and accurate reaction calculations consistent with nuclear structure (and ultimately with QCD) will most likely be provided by ``hybrid'' approaches that combine established few-body reaction techniques with EFT concepts, replacing \adhoc model assumptions with systematic, EFT-driven, inputs.
As an illustration, in the seminal work in Ref.~\cite{Capel:2018kss}, a halo EFT description of \isotope[11]{Be} was combined with the so-called dynamical eikonal approximation to calculate one-neutron transfer reactions of \isotope[11]{Be} on \isotope[208]{Pb} and \isotope[12]{C}.
While clearly only a limited set of atomic nuclei are amenable to a description within halo EFT, the technique is relevant for a number of reactions that are important for studies of exotic nuclei at rare-isotope facilities.
One might also envision broadening the approach to use more microscopic EFTs to describe the projectile, while still resorting ultimately to a few-body reaction picture.

In systems in which no experimental input is available to constrain LECs, halo EFT can be constrained using \textit{ab initio} results, provided such calculations can be performed.
This idea has been implemented, for example, in Ref.~\cite{Hagen:2013jqa}, where coupled-cluster calculations of \isotope[60]{Ca} were used to inform an EFT description of \isotope[60]{Ca}-$n$ scattering and to explore the possibility of an Efimov effect in \isotope[62]{C}.
For systems where experiments \emph{can} constrain the theory, replacing those inputs with more microscopic theory calculations (which in particular for heavy and exotic nuclei typically still feature large and/or unknown uncertainties) may not improve either the accuracy or the precision of predictions. 
However, from an intellectual point of view, it is certainly satisfying to follow a path like this and make systematic predictions from first principles, ``climbing'' the tower of theories depicted in Fig.~\ref{fig:tower} from bottom to top.

An alternative to combining different EFTs through the direct calculation of observables is provided by finite-volume simulations.
When quantum theories are formulated in a finite geometry, such as a cubic box with periodic boundary conditions, information about their physical properties is encoded in how the discrete energy spectrum depends on the size of the box, an important realization that goes back to the work of L\"uscher~\cite{Luscher:1990ux}.
At fixed box size one can envision simulating a nuclear halo state on the one hand using a microscopic description in terms of nucleons, using an interaction that is otherwise constrained, and on the one hand using a few-body halo EFT formulated in the same geometry.
The \apriori unknown parameters in the latter framework can then be determined by matching to the energy levels predicted by the microscopic description.
This procedure has great potential because it can be more informative than matching individual observables directly and because it is at the same time appealingly straightforward to implement.
It is also already being employed to determine certain LECs of pionless EFT from lattice QCD data~\cite{Sun:2022frr,Detmold:2023lwn}.
Calculations by the lattice EFT collaboration (see for example Refs.~\cite{Lee:2008fa,Elhatisari:2022qfr}), naturally performed in periodic boxes, can provide the needed microscopic energy levels, while efficient finite-volume few-body calculations are also being actively researched~\cite{Klos:2018sen,Konig:2020lzo,Bazak:2022mjh,Dietz:2021haj,Yapa:2022nnv,Konig:2022cya,Yu:2023ucq}.

\subsection{``The great nuclear simplification''}
\label{sec:Simplification}

An important feature of the halo/cluster EFT discussed above is its relative simplicity, achieved through a rather dramatic reduction of the number of dynamical degrees of freedom.
One might think that this is only possible owing to the peculiar structure of the systems it is tailored to describe, and that otherwise the majority of the nuclear chart, including rare isotopes without pronounced halo or cluster features, would have to be described by a more generic theory like chiral EFT (and even that may not provide a convergent expansion for the heaviest nuclei).
However, there are various indications from recent work that in a certain sense the nuclear interaction may be much ``simpler'' than one would naively think, such as:
\begin{enumerate}
\item The finding that the properties of (at least) light nuclei can be described by a \emph{perturbative} expansion around the unitarity limit (infinite S-wave nucleon-nucleon scattering lengths)~\cite{Konig:2016utl,Konig:2016iny,Konig:2019xxk}
 This expansion, which is constructed as a variant of pionless EFT, features a parameter-free two-nucleon interaction at leading order in addition to a three-nucleon force determined by a single three-nucleon datum.
 The actually finite values of the scattering lengths, along with other corrections and electromagnetic effects, enter subsequently at higher order treated in perturbation theory.
\item Work that characterizes the nuclear interactions with Gaussian potentials in order to capture the universal features of low-energy nuclear  physics~\cite{Gattobigio:2019omi,Deltuva:2020aws,Kievsky:2021ghz}.
 This approach has a resemblance to pionless EFT with a finite-range interaction: the range of the Gaussian potentials, tuned to reproduce observables, can be seen as fixed particular choice of ultraviolet (UV cutoff.
 However, by allowing for different ranges for different parts of the interaction, the approach is still phenomenological at its core and does not produce a systematic expansion with quantifiable uncertainties.
 Nevertheless, it is quite striking what level of accuracy can be achieved for few-nucleon system (or even nuclear matter~\cite{Kievsky:2018xsl}) in this manner, so overall this work provides a very interesting glimpse of what \emph{might} ultimately be achieved with a systematic but simple EFT approach.
\item The nuclear lattice EFT collaboration is pushing the idea of using a very simple nuclear interaction deep into the regime of medium-mass nuclei.
 Using an $SU(4)$-symmetric two-nucleon interaction, i.e., setting the scattering lengths in the two S-wave channels equal (and, of course, large), and adding contact three-nucleon interaction, Ref.~\cite{Lu:2018bat} produces a leading-order pionless EFT contact interaction (with a fixed cutoff determined by the lattice spacing).
 For the light mirror nuclei \isotope[3]{H} and \isotope[3]{He} this expansion had previously been shown to work well~\cite{Vanasse:2016umz}.
 By augmenting this with non-local ``smeared'' two-nucleon interaction, this work produces remarkable accurate results for various nuclear ground states up to mass number 50~\cite{Lu:2018bat}, and also for the spectrum of \isotope[12]C~\cite{Shen:2021kqr}.
 More recently, Ref.~\cite{Meissner:2023cvo} showed that also the monopole excitation in \isotope[4]{He} can be described with this simple interaction.
 While the smearing aspect is slightly difficult to relate to other formulations, at its heart this approach is again close to a non-perturbatively resummed variant of pionless EFT with an explicit finite range.
\item In this regard it is worth noting that resumming the effective range term that enters in pionless EFT at NLO has been found to give phenomenologically promising results~\cite{Lensky:2016djr,Bansal:2017pwn}.
\item In a similar spirit of simplifying the nuclear interaction to a minimum amount of necessary detail, Ref.~\cite{Fossez:2018gae} presents a study of neutron-rich helium isotopes.
 This approach starts with a \isotope[4]{He} core to which valence neutrons are coupled using a Woods-Saxon potential with parameters fit to reproduce low-energy $\alpha$-$n$ scattering.
 The interaction among valence nucleons in turn is provided by a finite-range potential acting only in spin-singlet channels.
 While an actual halo/cluster EFT would use only (regulated) contact forces, this approach is inspired by EFT in the sense that based on an analysis of the relevant scales of the problem it boils down the model to a ``leading order'' with few parameters.
 The accuracy achieved this way compared to available experimental data suggests again that indeed not much detail (such as explicit pion-exchange contributions) is needed to describe even exotic nuclear states far from stability.
\end{enumerate}

While at this point it is not at all clear what the final picture might look like, pursuing simplicity \emph{in a systematic way} may very well be the path that ultimately enables a consistent and rigorous description of nuclear structure and reactions.
Ultimately, the tower of EFTs shown in Fig.~\ref{fig:tower} may also inspire a reformulation of phenomenological models as, for instance, in Ref.~\cite{papenbrock11_1310,papenbrock15_1311} for deformed nuclei. 
In that regard, the construction of EFTs for the nuclear shell model and mean-field approaches will be major steps forward. 
However, it remains to be seen how the renormalization of nuclear forces in the medium, on which such approaches are based, can be formulated at the EFT level.

\subsection{Toward nuclear reactions from first principles}
\label{sec:omp}

As discussed in Sec.~\ref{sec:HaloEFT}, the EFT framework provides a rigorous way to describe nuclear reactions for systems that are amenable to a few-body description. 
However, in many reactions, such as knockout and transfer processes discussed in Sec.~\ref{sec:SRCandKO}, the structure of the target and possibly that of the projectile directly affect the dynamics. 
This situation generates a difficult many-body problem because, by definition, the extraction of reaction observables requires the explicit definition of all possible reaction channels, the number of which grows with the number of partitions of both the target and projectile, as well as with the number of states in each partition. 
In addition, the relative motion between clusters in each partition must be described in radial or momentum space. 
These issues constitute a significant part of the challenge of unifying the description of nuclear structure and reactions. 
Moreover, as the mass of the target increases, the density of low-lying states ``explodes,'' and soon enough only statistical descriptions remain as viable options. 
As will be discussed in Sec.~\ref{sec:Capture}, capture reactions, which are critical in many astrophysical processes, often fall in this category.

Approaches that treat all nucleons as active, such as the no-core shell model with continuum, have so far been limited to light nuclei~\cite{Deltuva:2008aa,Leidemann:2012hr,Navratil16_1956},
while less microscopic methods such as the Gamow shell model in the coupled-channel formalism can include a target with a core and reach higher masses~\cite{Jaganathen14_988,Fossez15_1119,Mercenne19_2262}.
However, in both cases only the two- or three-cluster relative motion can be treated explicitly.
In the near future, the use of symmetry-adapted approaches will extend the reach of \textit{ab initio} reactions involving two clusters into the medium-mass region before the statistical regime~\cite{Launey21_2578}.

A more economical avenue consists in reducing the many-body complexity into a few-body problem through the construction of so-called ``optical potentials'' that encode the projectile-target interaction (see Ref.~\cite{Hebborn:2022vzm} for a recent review).
In general, optical potentials include an imaginary part that represents the absorptive component accounting for processes not explicitly resolved in the calculation.
While such potentials can be purely phenomenological, recent developments aim at extracting them directly from \textit{ab initio} structure calculations. 
Many challenges remain in this direction, such as the lack of absorption due to the insufficient density of states given by many-body methods or the generalization of the approach beyond the two-body dynamics, 
but the general framework offers, in principle, a link between QCD and reaction observables~\cite{Holt:2022piv} when the underlying interaction is derived from an EFT. 
Due to the high sensitivity of reaction observables to thresholds, and to details of the structure in general, unresolved questions regarding both the construction of EFTs as well as their implementation in the many-body sector will need to be resolved to avoid an uncontrolled error propagation into the construction of optical potentials.

The development of few- and many-body methods including continuum couplings beyond one or two particles in the continuum is also important for the successful unification of structure and reactions. 
For instance, it is possible to provide benchmarks and inputs for reaction approaches by calculating energies, resonance widths, or asymptotic normalization coefficients (ANCs), 
but also wave functions that can be used for the construction of optical potentials including effects of continuum couplings within the target -- which has been shown to improve the problem of the insufficient absorption found within many-body calculations~\cite{Rotureau17_1953}.
For few-body systems, essentially exact calculations of many-body resonances are possible up to five particles~\cite{Lazauskas:2019hil,Yamaguchi23_2970}.
Finite-volume methods can also be used to study few-body resonances~\cite{Klos:2018sen,Dietz:2021haj,Yu:2023ucq}, complementing related approaches for bound 
states~\cite{Luscher:1985dn,Konig:2011xdn,Konig:2017krd,Yu:2022nzm} that can give access to ANCs and therefore also provide important inputs for direct capture calculations.
In order to improve theoretical calculations of resonances in many-body systems, a novel technique based on eigenvector continuation has recently been introduced in Ref.~\cite{Yapa:2023xyf}.

\subsection{Short-range correlations and knockout reactions}
\label{sec:SRCandKO}

Experimentally, direct reactions play a central role in studies of nuclear structure, as well as constituting a research topic unto themselves. They are uniquely powerful tools for probing the single-particle structure of nuclei through the overlap of initial and final quantum states in the involved species. 
However, a complete understanding of the dynamics of reactions is a challenge, and as such the model dependence of information extracted in reaction studies is a persistent limitation for the conclusions that can be drawn.

Of particular interest for RIB facilities that produce isotopes via in-flight fragmentation are intermediate-energy direct reactions, such as single-nucleon or two-nucleon knockout, typically performed at beam energies of $\sim$80-120~MeV/u. 
These reactions have been, and will continue to be, key tools for nuclear structure studies of the most exotic nuclei, providing access to information that allows the development of our understanding of shell structure and single-particle degrees of freedom. 
Akin to transfer reactions at lower energies, the cross-sections observed in knockout reactions on light nuclear (Be, C) targets populating specific final states relate to the occupancy of single-particle orbitals in the beam species. 
Information on the quantum numbers of the removed nucleon(s) is accessible through the width of the momentum distributions of the reaction residues~\cite{Hansen:2003dr}. 
However, despite the ubiquitous nature of direct reactions as an experimental tool, challenges remain.

For the vast majority of direct reactions (e.g. transfer, (e, e'p), knockout), there is a systematic reduction (or a suppression factor, $R_{S}$) of experimental cross-sections as compared to those calculated in the appropriate theoretical framework. 
For both transfer and (e, e'p) scattering the observed suppression is consistently $R_{S}$$\sim$ 0.5-0.6, a factor attributed to short-range correlations (SRC) between nucleons which are not captured in the low-momentum assumptions of the shell-model description of nuclei~\cite{Pandharipande:1997src}. 
However, for intermediate-energy nucleon knockout, the observed suppression is observed to be strongly correlated with the separation energy asymmetry ($\Delta S$), the difference in separation energies of the removed nucleon (proton or neutron) and the other species (neutron or proton)~\cite{Gade2008,Tostevin2014,Tostevin2021}. 
Efforts to connect observations from Jefferson Lab, which showed that the fraction of high-momentum protons increases in neutron-rich nuclei~\cite{Duer:2018SRC}, have demonstrated that the observed correlation between $\Delta S$ and $R_{S}$ in knockout can only be partially attributed to SRC and structural impacts~\cite{Paschalis2020}. 
It seems apparent that the reaction theory is playing a role in the measured systematic behaviour in knockout.

A concerted effort of theorists and experimentalists is required to get the most and most useful information out of knockout experiments and SRC studies. 
On the theory side it has been pointed out that, for example, spectroscopic factors are not actual observables because they run with the ``resolution scale'' of the interaction~\cite{Duguet:2014tua}, i.e., for a given interaction that describes the structure of a nucleus it is possible to perform a unitary transformation of the interaction that leaves all observables invariant, but would change spectroscopic factor defined as single-particle overlaps unless one defines them through an operator that is transformed consistently with the Hamiltonian. 
A consequence is that any two different interactions may give identical results for observables, but would generally yield inconsistent results for spectroscopic factors.
This analysis extends to SRCs, which can be characterized as ``scale and scheme dependent''~\cite{Furnstahl:2013dsa,Tropiano:2021qgf}, and moreover to optical potentials~\cite{Hisham:2022jzt}.
An important conclusion from these studies is that consistency is key to analyze knockout experiments and SRCs: the scale and scheme, for example in the form as a specific potential defining the interaction, need to be clearly specified and used consistently throughout the analysis.
For SRC studies, the so-called ``generalized contact formalism''~\cite{Weiss:2014gua,Weiss:2015pjw} has also emerged as a versatile theory framework for analyzing and interpreting experiments, providing guidance and inspiration for future work.

Looking forward on the experimental side, for direct reactions at fragmentation facilities to continue being the powerful tool that they have proven themselves to be, it is imperative that the theoretical description of the reactions continue to be refined and understood. 
The eikonal-model theory applied for the description of nucleon knockout includes consideration of the contributions from stripping (inelastic breakup), diffractive (elastic) breakup and Coulomb dissociation~\cite{Hansen:2003dr} and yet the correlation between $\Delta S$ and $R_{S}$ persists -- efforts are needed to attack this problem from both sides. 
Experimental work to identify cases which may enhance certain terms and thus provide insight into aspects of the reaction description will be key. 
Similarly, development of the reaction theory from first principles, even if limited to the lightest nuclei, may provide unique information to refine the eikonal description for application across the nuclear chart.

Future studies will also be moving to higher energy regimes and quasi-free scattering (QFS) reactions~\cite{Panin2021:QFS} with thick LH$_{2}$ targets will be essential tools for maximizing reaction luminosity in studies of the most exotic nuclei. 
QFS measurements are thus far typically considered at 350-450~MeV/u, and seem to show no evidence of correlation between $\Delta S$ and $R_{S}$~\cite{Atar2018:QFS} at these energies. 
Reconciling this situation with that of intermediate-energy nucleon knockout descriptions, and bridging the energy space between these two. remains a challenge which must be addressed as QFS-type reactions, or proton-induced knockout, become more common tools. 

\subsection{Capture reactions on rare isotopes}
\label{sec:Capture}

Direct reactions as described above only constitute a fraction of the plethora of processes that are interesting and relevant in nuclear physics. 
Another key class of reactions are capture processes which predominantly proceed via a compound nucleus. 
Charged particle, neutron, and gamma-capture reactions all play key roles in both applications (e.g., nuclear reactor modeling) as well as in astrophysical scenarios (as discussed in Sec.~\ref{sec:astro}). 
While the cross sections and dynamics of such reactions are critical information for a broad range of applications, the theoretical underpinnings of our understanding of these reactions are far from complete. 
Experimentally, there are significant challenges with performing direct measurements of capture reactions. Neutron-capture reactions are inherently difficult to measure due to the complexity of experiments with neutron beams and the lack of a neutron target. 
In exotic nuclei, low cross sections combined with the rates of radioactive ion beams make such measurements difficult even for charged-particle capture reactions.

Capture reactions can be separated into two categories, which are distinct both regarding their theoretical description and in terms of the experimental techniques needed to study them. 
The first is the case where the capture reaction populates individual states in the final compound nucleus (resonances). 
Both theoretically and experimentally the goal for such systems is to describe the properties of the individual states involved (energy, spin, parity) and, if at all possible, measure the resonance strength and potential interference with neighboring resonances.
When direct measurements of the resonance strength are not possible, then indirect techniques can be used. 
In this case the interplay between nuclear structure and nuclear reactions is more evident and important to understand. 

The second category collects processes where individual resonances are overlapping, in which case capture reactions need to be described within a statistical model. 
Within such a model, the interaction between the incoming particle and the nucleus is described through an optical model potential, as discussed in Sec.~\ref{sec:omp}. 
In addition, the nucleus is described using statistical properties like the nuclear level density (NLD) and the spin distribution, while the de-excitation of the nucleus is described through a $\gamma$-ray strength function ($\gamma$SF). 
Each of these quantities is calculated through different theoretical models (see for example a recent review~\cite{Lar19}), which are optimized using experimental data along the valley of stability. 
Naturally, when moving away from stability, the model predictions diverge and therefore the theoretical uncertainties in capture cross sections can reach up to two orders of magnitude. 
Experimentally there are very few measurements of charged-particle capture reactions on unstable nuclei, and even fewer for neutron capture on long-lived radioisotopes. 
Therefore, indirect techniques are needed to provide constraints to the theoretical models. 
As discussed in Sec.~\ref{sec:astro}, indirect techniques for neutron-capture reactions are heavily used in astrophysical processes far from stability~\cite{Lar19}.

\section{FROM NUCLEAR PROPERTIES TO ASTROPHYSICAL PHENOMENA}
\label{sec:astro}

\subsection{Overview of processes in nuclear astrophysics}

The connection to astrophysical phenomena has been realized and investigated since the early days of the field of nuclear physics. 
Nuclear reactions were identified as the source of energy generation in our Sun and other stars, and quickly this led to the conclusion that the same nuclear reactions could be responsible for synthesizing new elements. 
Still, in the early 1950s it was unclear whether heavy elements were produced already during the Big Bang or whether there was a possibility that the stars themselves were synthesizing them. 
The debate was resolved in 1952 when astrophysicist Paul Merrill~\cite{Merrill1952} observed technetium (Tc) lines in a stellar spectrum. 
With the longest isotope of Tc having a half-life of 4.2 My, the observed Tc could not have come from the Big Bang. 
This breakthrough observation led to an exploration of all astrophysical processes that could create the known chemical elements, which was published in 1957~\cite{B2FH}. 
The conclusions of this seminal work are largely still valid today, although a lot has changed regarding the details in the six decades since this early work.

The lightest of the elements, H, He, and Li, are all produced mainly during the Big Bang. 
The nuclear reactions that drive their synthesis involve mostly stable isotopes and have been studied extensively, although higher accuracy is still needed~\cite{Cyb16}. 
All other elements are formed in stars. 
Up to the region of iron, new elements are formed mainly in the quiescent phase of a star's life, although burning cycles that help balance a star's gravitational collapse while at the same time producing heavier elements. 
The nuclear reactions involved in stellar burning continue to be investigated today; however, the majority of these reactions involve stable nuclei and are thus outside the scope of the present review article. 

\begin{figure}[h]
\includegraphics[width=4in]{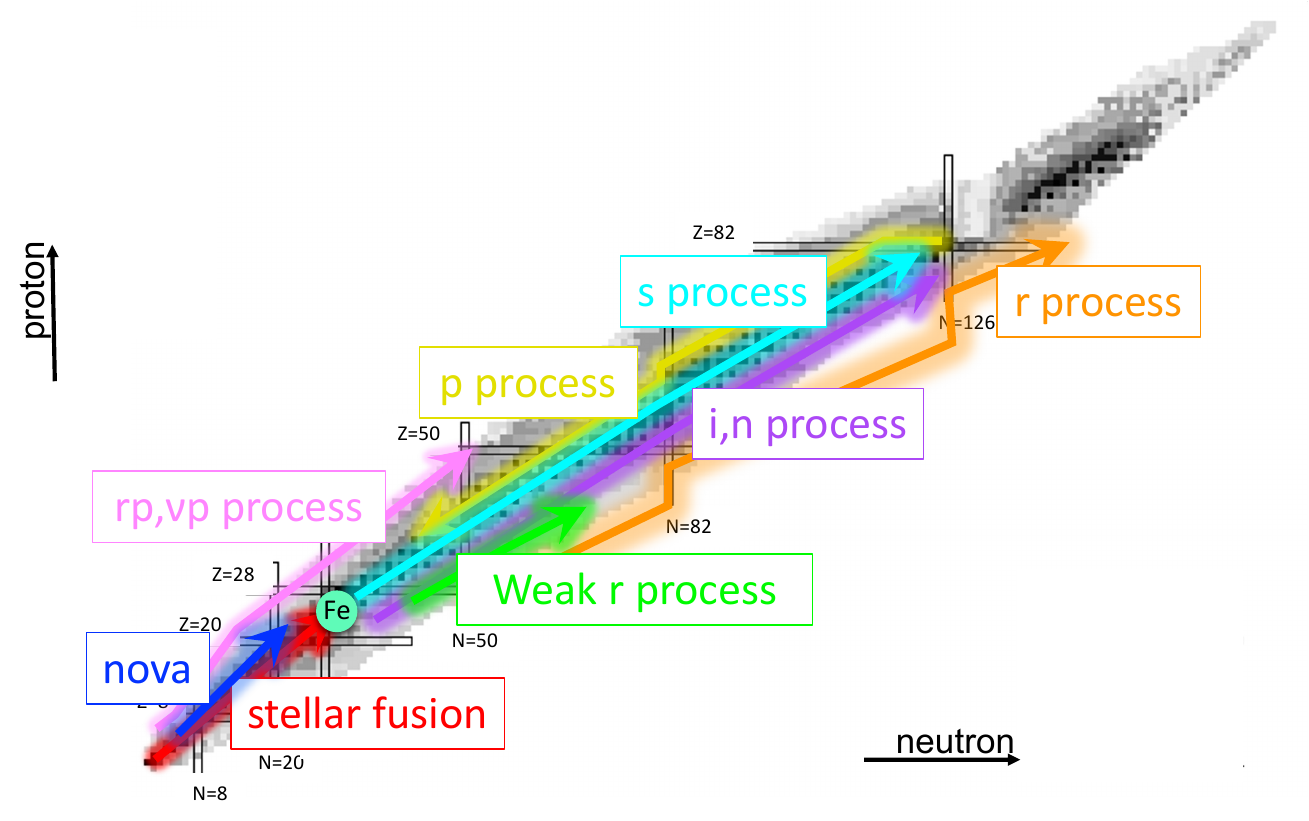}
\caption{Chart of nuclei with the astrophysical processes marked roughly in the regions they are flowing through. }
\label{fig:astro-chart}
\end{figure}

Heavy element nucleosynthesis is far more complex. 
Up until two decades ago the available observables could be mostly explained by three nucleosynthesis processes. 
Two of them involve the capture of neutrons followed by $\beta$ decay, through the slow (s)~\cite{Kap11} and rapid (r)~\cite{Cow21} processes. 
A third process (p process) was introduced to explain the production of roughly 35 neutron-deficient isotopes that cannot be produced by the other two processes~\cite{Rau13}. 
The picture became a lot more complicated when more and more astronomical observations revealed abundance patterns that could not be explained by the three aforementioned processes. 
Therefore, the need for additional nucleosynthesis processes led to a revived interest in the community to identify, constrain and validate realistic scenarios that reproduce the new observations. 
Some of the proposed scenarios include the $\nu p$ process~\cite{Fro06} in core-collapse supernova, the Light Element Primary Process (LEPP)~\cite{Arc11}, the i process~\cite{Her11}, the n process~\cite{Bla76} and the weak r process~\cite{Ish05}. 
Therefore, the current and future goal of heavy element nucleosynthesis is to reduce the uncertainties associated with each of these processes and identify if and how much they contribute to the observed abundances. 

On top of stellar nucleosynthesis, the field of nuclear astrophysics also aims to understand extreme astrophysical phenomena. 
These include novae, supernovae, X-ray bursts and neutron-stars. 
Instead of abundances, the observables may be light curves at different wavelengths, neutrinos, $\gamma$-ray observations from long-lived radioisotopes, and gravitational waves. 
Similar to nucleosynthesis processes, an accurate understanding of the stellar conditions as well as the properties of the involved nuclei is critical for reproducing the observables.

From the nuclear physics point of view, each stellar process and phenomenon requires a different set of nuclear inputs. 
Nuclear physicists work closely with astrophysics modelers to identify the important properties and (a) measure them in the lab directly, if possible, (b) provide indirect experimental constraints, or (c) perform theoretical calculations. 
In the present review, our goal is not to describe all possible processes and nuclear inputs, but rather to focus on some concrete examples that rely on input from rare isotope facilities. 

\subsection{Current status of nuclear astrophysics with rare isotopes}

\paragraph*{Nucleosynthesis}
One of the least known processes in nuclear astrophysics is the r process, despite the fact that it is considered responsible for the synthesis of about half of the isotopes of heavy elements. 
One of the main difficulties in modeling the r process accurately comes from the nuclear input. 
The involved nuclei are very exotic, which makes it challenging to study experimentally, and theoretical predictions are not well constrained. 
To date most of the relevant isotopes are not available for experiments at radioactive beam facilities, although next-generation facilities such as FRIB have started operation and will ultimately give access to a large fraction of the important nuclei. 
The nuclear properties that affect r-process calculations include nuclear masses, $\beta$-decay properties, neutron-capture rates, fission properties, isomers, and more. 
A review of rare-isotope facilities and their connection to r-process nucleosynthesis was published recently~\cite{Horowitz.2019}. 
Here we limit ourselves to a high-level summary and some specific examples. 

Most of the important nuclear properties can be directly measured, as long as suitable rare isotope beams are available. 
In this case the goal is to develop more powerful facilities that can provide the isotopes of interest. 
This is where next generation rare isotope facilities, like FRIB, will have a direct and major impact. 
When beams are available, mass measurements, $\gamma$-ray spectroscopy, reaction measurements and decay studies all contribute to the study of nuclear structure and how it evolves far from stability. 
Traditional magic numbers, and new emerging ones, both have a direct impact on the r-process flow and the final abundance patterns. 
Therefore, nuclear structure studies of unexplored rare isotopes, such as halo and cluster states as discussed in Sec.~\ref{sec:HaloEFT}, are critical.

Unfortunately, even with full access to the relevant isotopes, neutron-capture rates on short-lived nuclei are currently extremely challenging to measure directly. 
Therefore, indirect approaches are needed to constrain them. 
These techniques offer a unique opportunity for connections between nuclear structure, nuclear reactions, nuclear theory, nuclear experiment and astrophysics. 
Theoretical approaches for capture reactions are discussed in Sec.~\ref{sec:Capture}.

Finally, a major aspect of the r process, and probably the furthest from experimental reach, is nuclear fission. 
In this case, theoretical calculations, validated on existing experimental data, are essential to provide fission properties in r-process models. 
However, further discussion of this is beyond the scope of this article and we thus refer the interested reader to dedicated reviews such as Ref.~\cite{Bender:2020oze}.

A group of processes that generally occur in conditions between the s and r processes have been proposed to reproduce ``strange’’ astronomical observations. 
Although these processes all take place in different astrophysical environments, neutron densities and time scales, we group them here together because the nuclear physics needs to understand them are similar. 
Since these processes involve nuclei that are only a few neutrons away from stability, most of the nuclear physics properties are known experimentally. 
The most significant uncertainties come from the unknown neutron-capture rates (see Sec.~\ref{sec:Capture}). 
A reduction in the nuclear uncertainties will help identify the conditions under which these processes can create the observed abundance patterns, and also the possible contributions to solar system or other abundances. 

\paragraph*{Extreme astrophysical events}
In the era of multi-messenger astronomy, signals from stellar explosions provide insights into extreme astrophysical conditions. 
X-ray observations from accreting neutron stars (X-ray bursts) provide a window into the properties of matter at extreme densities inside neutron stars.
Neutrinos and electromagnetic observations from nearby supernovae inform about the elusive explosion mechanism. 
$\gamma$-ray measurements and stardust grains from novae explosions help us understand better the nucleosynthesis and explosion physics of these sites. 
Gravitational waves and electromagnetic signatures let us probe neutron-star mergers. 

With such a rich collection of observables and sites, the nuclear physics needs are also quite diverse. 
A major success of the nuclear astrophysics community is the fact that all reactions involved in nova nucleosynthesis are now, to a large extent, measured experimentally. 
This allows for more accurate modeling of this stellar event, and better estimates of the contributions that nova explosions have to galactic nucleosynthesis. 

In core collapse supernova (CCSN), one of the main nuclear physics inputs is weak reaction processes, in particular electron capture (EC) rates. 
ECs regulate the electron density and strongly influence the dynamics of the collapse. 
They also produce the neutrinos that carry energy out of the collapsing core. 
In the lab, electron captures can only be measured within the EC Q-value. However, significant experimental effort has been dedicated to studying the EC process indirectly through charge-exchange reactions~\cite{Lan21}. 
Currently we are far from having experimental constraints on all relevant EC rates; therefore, this is another case where astrophysical models have to rely heavily on theoretical calculations. 
Significant effort is devoted to constraining the nuclear theory with experimental data where possible. 
Another key nuclear input related to CCSN is the nuclear reaction network that produces key radioisotopes observed by $\gamma$-ray telescopes. 
\isotope[26]{Al} and \isotope[60]{Fe} are two of the dominant radioisotopes, and significant effort has been devoted across the community to measuring relevant reactions and decay properties to characterize their production and emission during CCSN events. 

Accreting neutron stars provide a unique insight into dense-matter physics. 
One key observable is type I X-ray bursts, which are powered by thermonuclear explosions and have recurrence times of hours to days. 
With more than 100 such systems known in our galaxy, these are the most frequently observed explosions and provide a rich and high-precision dataset. 
The modeling of these events, and the conclusions from these models on dense matter properties, are hindered by unknown nuclear reactions that drive the thermonuclear explosions. 
Sensitivity studies have identified important reactions~\cite{Cyb16}, but to date only a small number has been constrained experimentally since the relevant cross sections are small and the available beam rates not yet high enough.
Indirect approaches are also used here (similar to the neutron-capture reactions mentioned in Sec.~\ref{sec:Capture}). 
In addition, the possible mechanism that cools the neutron star crust was identified recently~\cite{Sch14} as the cycle of alternating electron captures and $\beta^-$ decays (Urca process). 
The nuclear physics uncertainties associated with this process are significant and so far hinder our ability to accurately describe this potential cooling mechanism. 

The Urca process is dominated by pairs of isotopes for which the $\beta^-$ decay predominantly feeds the ground-state or low-lying excited state of the final nucleus, followed by an electron capture. 
Since the majority of the involved nuclei are far from stability, new mass measurements are needed for an accurate estimate of the $\beta$-decay Q-value. 
In addition, significant effort was dedicated to $\beta$-decay measurements of the ground-state to ground-state feeding intensity, in order to identify viable Urca pairs~\cite{Ong20}.

\subsection{Nuclear structure for astrophysics}

From the early days of nuclear astrophysics it was clear that the details of nuclear structure are directly linked to stellar processes and the observables we have from the universe. 
The high abundance of Fe-peak nuclei is linked to their high per-nucleon binding energy; the location of the s-process and r-process abundance peaks is linked to neutron magic numbers at N=50, 82 and 126; the triple-alpha process would not be nearly efficient enough without the presence of a resonance in \isotope[12]{C} (Hoyle state). 
These are just a few examples of direct connections between a nuclear structure property and an astronomical observable. 
It is therefore clear that without an accurate knowledge of nuclear structure, and especially how it evolves far from stability, we cannot hope to have an accurate description of astrophysical processes. 

During the last few decades several new phenomena have been observed when studying nuclei at extreme neutron-to-proton ratios, as described in Sec.~\ref{sec:exoticExperiment}. 
These phenomena, and in particular shell evolution, can have direct impacts on the final abundance patterns that are observed today. 
One example of such an impact was shown in Ref.~\cite{Lor15}, where new half-life measurements at the RIKEN facility resulted in a better reproduction of the solar r-process abundance pattern. 
Similar measurements at FRIB in the coming years are highly anticipated, in an effort to disentangle the contributions from various astrophysical processes. 

Despite the experimental efforts and new facilities, there is still a significant number of nuclei that cannot be reached by experiment. 
It is therefore also of major importance to ensure that the available theoretical calculations are as reliable as possible. 
For this reason, effort is dedicated in the community (and will continue to be) into measurements that are not necessarily the most important ones from the astrophysical point of view, but which can be used to test theoretical predictions. 
One example is the measurement of $\beta$-decay properties, such as half-lives~\cite{Lor15}, $\beta$-delayed neutron emission probabilities~\cite{Pho22} and $\beta$-decay intensity distributions~\cite{Dom21}.

\subsection{Reactions for astrophysics}

Nuclear reactions play a major role in astrophysical processes since they drive the energy release and absorption, as well as the element synthesis and destruction. 
For stellar processes that evolve around the valley of stability many of the relevant reactions have been measured experimentally. 
Still, many continue to be elusive since their cross sections at the relevant stellar temperatures are extremely small. 
In that regard, underground facilities have made (and continue to make) significant progress in measuring important reactions in low-background environments~\cite{Bro18}. 
The picture is very different when looking at astrophysical processes that take place even a few steps from stability. 
In these cases, radioactive beams are necessary for performing these measurements, and these are typically available with low intensities or not at all. 

Radiative capture reactions (described in Sec.~\ref{sec:Capture}) often dominate astrophysical processes. 
Recoil separators have been developed for this reaction category, with successful radioactive beam measurements~\cite{Rui14}. 
Especially for light masses, the reaction rate is dominated by a small number of strong resonances, and the focus of previous measurements has been on identifying and if possible directly measuring the resonance strengths. 
Moving to heavier masses, individual resonances cannot be resolved, and the reaction rate varies effectively smoothly as a function of energy. 
However, due to the higher Coulomb barrier in heavier nuclei, the capture reaction cross sections become smaller, and in addition the resolving power of recoil separators gets worse. 
Three additional techniques have been developed recently for higher mass capture reaction measurements. 
The $\gamma$-summing technique in inverse kinematics focuses on the measurement of the emitted $\gamma$ rays, therefore the recoil-beam separation does not matter. 
This technique uses large-volume $\gamma$-ray detectors, significantly increasing the detection efficiency of the setup. 
The $\gamma$-summing technique has been successfully applied to stable beam reactions~\cite{Palmisano.2022,Tsantiri.2023} and the first radioactive beam experiment was recently completed at FRIB. 
The use of a storage ring for capture reaction measurements was also successfully demonstrated recently with stable beams at GSI~\cite{Glo19}. 
Storage rings have the advantage of circulating the beam up to a million times per second, effectively increasing the beam intensity that interacts with the target.
The only heavy-mass radioactive-beam proton capture reaction ever measured is the \isotope[83]{Rb}(p,$\gamma$)\isotope[84]{Sr}~\cite{Lotay.2021} done at TRIUMF. 
A combination of high-resolution $\gamma$ detection and recoil measurement allowed for the successful identification of the reaction products and the extraction of the cross section at astrophysical energies. 
Despite the new developments, we are still far from being able to measure all relevant capture reactions directly, therefore indirect techniques, like the ones mentioned in section~\ref{sec:Capture} are still essential. 
Looking towards the future, the increased beam intensities of the next generation rare-isotope facilities will allow for more radiative capture reactions with radioactive beams, either directly or using indirect techniques. 
 
($\alpha$,p), ($\alpha$,n) and (p,n) reactions were also identified as important drivers of particular astrophysical processes~\cite{Bli17,Cyb16,Fro06}. These reactions often have higher cross sections and successful direct measurements with stable and radioactive beams have been performed. Different techniques have been developed for measuring these reactions with radioactive ion beams. One such technique is the use of a recoil separator. Recoil separators are traditionally used for capture reaction measurements, however, the new generation of separators (e.g., SECAR at FRIB) have the necessary resolving power and can be used for the this new category of reactions as well. Another new approach is the use of a multi sampling ionization chamber (MUSIC)~\cite{Avi17}.
These active-target detection systems allow for the simultaneous measurement of a broad range of excitation energies, and also multiple reaction channels. The use of MUSIC for this purpose has been established at Argonne National Lab and has also been applied at FRIB and other facilities. Future measurements with these and other techniques will benefit greatly from the increased beam intensities that the next generation rare isotope facilities will offer. 

\subsection{Plasma effects}

As mentioned already, nuclear structure and reactions measurements provide key inputs to understanding astrophysical phenomena, but in many cases it remains a challenge to access directly the information of interest under the conditions relevant in astrophysical sites. 
This includes the difficulty of performing direct measurements at the appropriate energies, but also to take into account the potential impact of the plasma-like environment in stellar systems. 
The latter may modify decay (electron-capture) rates or reaction rates with additional (induced) atomic processes, complicating the delicate equilibrium of the system.

To date, there has been only limited direct experimental effort to couple nuclear physics measurements with the plasma physics required to capture this complexity. 
The \isotope[3]{He}(d,p)\isotope[4]{He} fusion reaction was explored in a plasma environment at temperatures of a few keV using \isotope[3]{He} atoms and the interaction of ultra-fast laser pulses with molecular deuterium clusters~\cite{BarbuiBang.2013}. 
This reaction measurement followed years of work in the domain of plasma physics to explore and understand the properties of plasmas formed in laser-induced explosion of deuterium clusters (see, for example, Refs.~\cite{Ditmire.1999,Krainov.2002,BangBarbui.2013}). 
However, such efforts have thus far been limited to light-ion fusion reactions. 
Work to understand the impact of the plasma environment on reactions with heavier nuclei or on decay properties has been constrained thus far to theory.

Plasma physics and high-power ultra-fast laser technology have been advancing rapidly, in parallel with radioactive ion beam facilities and capabilities. 
Looking forward, it seems clear that these research areas will be able to come together to further experimental efforts to understand key decay and reaction properties in plasma environments more relevant to the astrophysical sites. 
However, the challenges are substantial -- in addition to the difficulty of coupling ion beam facilities with the technology (e.g., laser systems) required to create plasma conditions, development will also be required for diagnostic and experimental detection systems. 
This is not an effort that will be completed in short-order, but does represent an important future direction for a more complete understanding of nuclear physics in astrophysical environments.

\section{CONCLUSIONS}

The science of rare isotopes is situated at the intersection between nuclear structure, nuclear reactions, and processes relevant for nuclear astrophysics. 
New facilities, including FRIB, which have begun operation or will do so within the next decade, will enable a host of new measurements and the exploration of a large part of the \textit{terra incognita} located between the valley of stability and the neutron drip line. 
In addition to advances in detectors and measurement techniques, the scientific programs at these facilities will require critical input from nuclear theory to properly guide, analyze, and interpret experiments involving exotic nuclei. 

In this article, we attempted to highlight some of the exciting new phenomena observed far from the valley of stability, with particular emphasis on exploring the connections and interplay between nuclear structure and reactions, both from theoretical and experimental perspectives.
In Sec.~\ref{sec:emergent-phenomena}, we reviewed selected phenomena mostly observed in light nuclei near the drip lines and speculated over what new phenomena could appear in heavier nuclei and the challenges they will pose for their coming experimental investigations and theoretical descriptions. 

In Sec.~\ref{sec:EFT}, we discussed how nuclear EFTs have come a long way since their inception and are widely celebrated for connecting nuclear physics to QCD.
While the study of light nuclei already opened important and interesting questions regarding the construction and practical implementation of nuclear EFTs, new observations of exotic nuclei, possibly having new emergent scales, will bring additional challenges and opportunities.
There is mounting evidence, however, that EFT concepts may be used to construct ``simple'' interactions that focus exactly on what is essential for the description of nuclei, including exotic states.
In Sec.~\ref{sec:Simplification}, we collected an overview of some of this work, putting it into the larger perspective, with the hope of inspiring further research along these lines.

Nuclear structure and reactions in exotic nuclei have a direct impact on our understanding of stellar processes. In Sec.~\ref{sec:astro}, we gave an overview of the present status of the field, with an emphasis on the processes in which rare isotopes play a vital role. The long history of the field, and in particular the last decade of discoveries both in astronomical observations and in the science of rare isotopes, have shown that nucleosynthesis is more complex than previously thought. Looking forward, the field will attempt to disentangle the contributions of different astrophysical processes to the synthesis of elements in the Universe. This can only be achieved with an accurate description of rare isotopes.

In this article, we have highlighted specific processes, such as knockout and related reactions (Sec.~\ref{sec:SRCandKO}), where close connections between experiment and theory are particularly important. A unified treatment of nuclear structure and reactions is key for rare isotopes and exotic nuclei,
as discussed in Sec.~\ref{sec:omp} and more specifically in Sec.~\ref{sec:Capture} for capture reactions. 

However, more broadly, truly understanding the physics of rare isotopes and using these systems to improve our knowledge of the nuclear interaction as well as that of astrophysical phenomena requires close collaboration and exchange between theorists and experimentalists. This is a challenge that we believe our field is ready to meet, and we look forward to the next decades of results which will continue to expand our vision of rare isotopes. 

\section*{DISCLOSURE STATEMENT}
The authors are not aware of any affiliations, memberships, funding, or financial holdings that might be perceived as affecting the objectivity of this review.

\section*{ACKNOWLEDGMENTS}

We thank F.~Nunes for bringing the authors together and suggesting this review.
The work of S.K.\ is supported by the U.S.\ NSF (PHY-2044632) and by the U.S.\ DOE (DE-SC0024520 -- STREAMLINE Collaboration and DE-SC0024622).
The work of K.F.\ is supported by the U.S.\ NSF (PHY-2238752) and the U.S.\ DOE (DE-SC0024646 -- STREAMLINE Collaboration).
This material is based upon work supported by the U.S.\ DOE, Office of Science, Office of Nuclear Physics, under the FRIB Theory Alliance, award DE-SC0013617.
The work of A.S.\ is supported by the U.S.\ NSF (PHY 2209429).
The work of H.L.C. is supported by the U.S.\ DOE, Office of Science, Office of Nuclear Physics, under Contract No.\ DE-AC02-05CH11231 (LBNL) and support under the DOE Early Career Research Program.


\bibliographystyle{ar-style5.bst}

\end{document}